\documentclass[aip,reprint,showpacs,twocolumn,superscriptaddress,floatfix]{revtex4-1}
\usepackage{epsfig}
\usepackage[T1]{fontenc}
\usepackage[utf8]{inputenc}
\usepackage{amsmath}
\usepackage{amssymb}
\usepackage{amsfonts}
\usepackage{mathptmx}
\usepackage{dcolumn}
\usepackage{eucal}
\usepackage{bm}
\usepackage{color}
\usepackage[colorlinks,linkcolor=blue,citecolor=blue]{hyperref}

\usepackage{epstopdf}
\begin{document}



\title{A normal metal tunnel-junction heat diode}

\author{Antonio Fornieri}
\email{antonio.fornieri@sns.it}
\affiliation{NEST, Istituto Nanoscienze-CNR and Scuola Normale Superiore, I-56127 Pisa, Italy}

\author{María José Martínez-Pérez}
\affiliation{NEST, Istituto Nanoscienze-CNR and Scuola Normale Superiore, I-56127 Pisa, Italy}

\author{Francesco Giazotto}
\email{giazotto@sns.it}
\affiliation{NEST, Istituto Nanoscienze-CNR and Scuola Normale Superiore, I-56127 Pisa, Italy}




\begin{abstract}
We propose a low-temperature thermal rectifier consisting of a chain of three tunnel-coupled normal metal electrodes. We show that a large heat rectification is achievable if the thermal symmetry of the structure is broken and the central island can release energy to the phonon bath. The performance of the device is theoretically analyzed and, under the appropriate conditions, temperature differences up to $\sim$ 200 mK between the forward and reverse thermal bias configurations are obtained below 1 K, corresponding to a rectification ratio $\mathcal{R} \sim$ 2000. The simplicity intrinsic to its design joined with the insensitivity to magnetic fields make our device potentially attractive as a fundamental building block in solid-state thermal nanocircuits and in general-purpose cryogenic electronic applications requiring energy management. 
\end{abstract}


\maketitle


The evolution of modern electronics has been boosted by the introduction of non-linear elements like diodes and transistors. The latter represented fundamental milestones for the control of electric currents and for the execution of logic operations in contemporary electronic devices. On the other hand, in the last decade we witnessed an explosion of interest in the investigation of thermal transport at the nanoscale. \cite{GiazottoRev,Dubi,LiRev,BenentiRev} Although appearing more difficult to control, heat transfer in solids embodies one of the most interesting and promising topics in nanoscience. Yet, the implementation of solid-state circuits that could enable the manipulation of heat currents is still at the earliest stage of development. These structures would be at the basis of emerging fields such as nanophononics,\cite{LiRev} thermal logic \cite{LiRev} and coherent caloritronics.\cite{GiazottoNature,MartinezNature,MartinezRev} Moreover, they would be of great conceptual and technological interest for many other research fields such as solid-state cooling,\cite{GiazottoRev,GiazottoTaddei} ultrasensitive cryogenic radiation detection \cite{GiazottoRev,GiazottoHeikkila} and quantum information.\cite{NielsenChuang,SpillaArxiv} In this context, the realization of highly-efficient thermal diodes, i. e., devices in which thermal transport along a specific direction is dependent upon the sign of the temperature gradient,\cite{Starr,RobertsRev} appears as a crucial first step.

So far, several proposals have been made to design thermal rectifiers dealing, for instance, with phonons,\cite{Terraneo,Li,Segal,Segal2} electrons \cite{GiazottoBergeret,MartinezAPL,Ren,Ruokola1,Kuo,Ruokola2,Chen} and photons.\cite{BenAbdallah} On the experimental side, encouraging results were obtained in devices that exploited phononic \cite{Chang,Kobayashi,Tian} or electronic \cite{Scheibner} thermal currents. In the latter case, a large rectification effectiveness has been demonstrated very recently in a hybrid superconducting device at cryogenic temperatures.\cite{MartinezArxiv}

In this Letter we propose and theoretically analyze the performance of a thermal diode consisting of three normal metal islands (N$_1$, N$_2$, and N$_3$) connected by two thin insulating layers (I), thereby forming a N$_1$IN$_2$IN$_3$ junction. As we shall argue, this simple design can offer outstanding rectification effectiveness at low temperatures (below 1 K), provided that two conditions are satisfied: (i) the thermal symmetry of the system must be broken, and (ii) N$_2$ must be coupled to the phonon bath. The proposed device could be easily realized with conventional nanofabrication techniques \cite{GiazottoNature,MartinezNature,MartinezArxiv} and could be immediately exploited in low-temperature solid-state thermal circuits. Moreover, it would be virtually unaffected by magnetic fields ensuring high performance also in conditions where a hybrid superconducting thermal diode \cite{MartinezArxiv} could lose its effectiveness.
\begin{figure}[t]
\centering
\includegraphics[width=1\columnwidth]{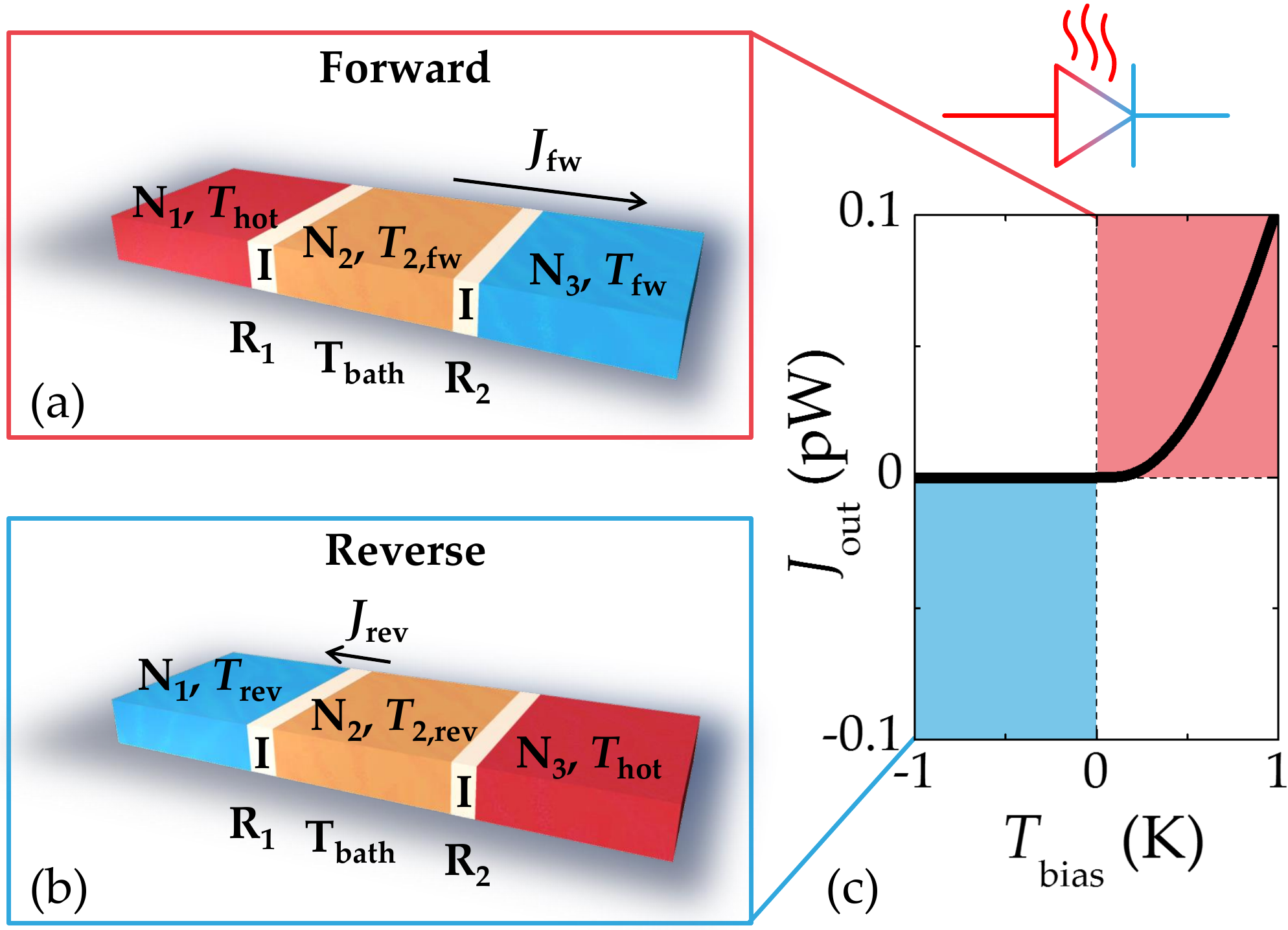}
\caption{ Panels (a) and (b) show respectively the forward and reverse thermal bias configurations of the proposed heat current rectifier. The device is made of three normal metal islands (N$_1$, N$_2$, and N$_3$) connected by two thin insulating layers (I), thereby forming a N$_1$IN$_2$IN$_3$ junction with a total in-series resistance of $R_1 +R_2$. 
(c) Output heat current $J_{\rm out}$ vs. bias temperature $T_{\rm bias}$ of a thermal diode with $\mathcal{R}\gg 1$ [calculated for the design shown in Fig. \ref{Fig2}(d)]. The red and blue quadrants correspond to the forward and reverse thermal bias configurations, respectively.}
\label{Fig1}
\end{figure} 
We shall start, first of all, by defining two parameters that will help us to describe the diode's performance, i.e., the rectification effectiveness $\mathcal{R}$ and the thermal efficiency $\eta$. Toward this end, we analyze in more detail the structure of our system. As shown in Fig. \ref{Fig1}(a) and \ref{Fig1}(b), N$_2$ is coupled to N$_1$ and to N$_3$ by means of two tunnel junctions characterized by resistances $R_1$ and $R_2$, respectively. For simplicity, from now on, we will set $R_1= 500$ $\Omega$ and consider only the parameter $r=(R_2/R_1)\geq 1$, accounting for the asymmetry of the device. N$_1$ and N$_3$ act as thermal reservoirs and are used to establish a temperature gradient across the device. Electrode N$_2$, instead, represents the core of the diode, since it controls the heat flow from a reservoir to the other by releasing energy to the phonon bath (residing at temperature $T_{\rm bath}$). 
In the \emph{forward} configuration, the electronic temperature of N$_1$ is set to $T_{\rm hot}> T_{\rm bath}$. This temperature bias leads to heat currents $J_{\rm in,fw}$ and $J_{\rm fw}$ flowing into N$_2$ and N$_3$, respectively. On the other hand, in the \emph{reverse} configuration the electronic temperature of N$_3$ is set to $T_{\rm hot}$, generating the heat currents $J_{\rm in,rev}$ and $J_{\rm rev}$ flowing into N$_2$ and N$_1$, respectively. Under these assumptions, we can define the rectification effectiveness as:
\begin{equation}
\mathcal{R}=\frac{J_{\rm fw}}{J_{\rm rev}}.
\end{equation}
In general, a highly-effective thermal diode is characterized by $\mathcal{R}\gg 1$ or $\ll 1$. In our case, since we have chosen $r\geq 1$, the forward configuration results to be the most transmissive, as we will show. The thermal diode's response can be heuristically compared to that of the well-known electric diode by plotting the output current $J_{\rm out}$ vs. the bias temperature $T_{\rm bias}$. In the forward configuration we define $T_{\rm bias}= T_{\rm hot}-T_{\rm bath}$ and  $J_{\rm out}= J_{fw}$, whereas in the reverse configuration $T_{\rm bias}=-(T_{\rm hot}-T_{\rm bath})$ and $J_{\rm out}= -J_{rev}$.
As shown in Fig. \ref{Fig1}(c), if $\mathcal{R}\gg 1$ the behavior of $J_{\rm out}$ vs. $T_{\rm bias}$ is strongly asymmetrical, indicating a drastic mismatch of the diode's heat transport properties between the forward and the reverse configuration. Yet, another important parameter that has to be considered is the thermal efficiency. In the transmissive configuration, it can be defined as:
\begin{equation}
\eta=\frac{J_{\rm fw}}{J_{\rm in,fw}},
\end{equation}
indicating the fraction of power that is transferred from N$_1$ to N$_3$. An ideal diode should exhibit $J_{\rm fw}=J_{\rm in,fw}$ and $J_{\rm rev}=0$, leading to $\mathcal{R}\rightarrow \infty$ and $\eta=1$.

We describe now the equations governing heat transport in our device. First, if we consider two N electrodes residing at electronic temperatures $T_1$ and $T_2$ (with $T_1\geq T_2$ for definiteness) coupled by means of a tunnel junction, the stationary electronic thermal current flowing through the junction can be written as:\cite{GiazottoBergeret} 
\begin{equation}
J_\mathrm{e}(T_1,T_2)=\frac{k_{\mathrm{B}}^2 \pi^2}{6e^2 R_\mathrm{N}}(T_1^2-T_2^2),\label{Qe}
\end{equation}
where $R_\mathrm{N}$ is the contact resistance, $e$ is the electron charge and $k_\mathrm{B}$ is the Boltzmann's constant. Moreover, we must take into account the heat exchanged by electrons in the metal with lattice phonons:\cite{MartinezNature,Maasilta}
\begin{equation}
J_{\mathrm{e-ph}}(T,T_\mathrm{bath})=\Sigma \mathcal{V} (T^n-T_\mathrm{bath}^n).\label{Qeph}
\end{equation}
Here $\Sigma$ is the material-dependent electron-phonon coupling constant, $\mathcal{V}$ is the volume of the electrode and $n$ is the characteristic exponent of the material. In this work we will consider two materials that are commonly exploited to realize N electrodes in nanostructures, i.e., copper (Cu) and manganese-doped aluminum (AlMn). The former is typically characterized by $\Sigma_{\rm Cu}=3\times 10^9$ WK$^{-5}$m$^{-3}$ and $n_{\rm Cu}=5$,\cite{GiazottoRev,GiazottoNature} while the latter exhibits $\Sigma_{\rm AlMn}=4\times 10^9$ WK$^{-6}$m$^{-3}$ and $n_{\rm AlMn}=6$.\cite{MartinezNature,MartinezArxiv} Furthermore, we assume that N$_1$ and N$_3$ are identical, with volumes $\mathcal{V}_1=\mathcal{V}_3= 2 \times 10^{-20}$ m$^{-3}$.\cite{MartinezArxiv}
\begin{figure}[t]
\centering
\includegraphics[width=1\columnwidth]{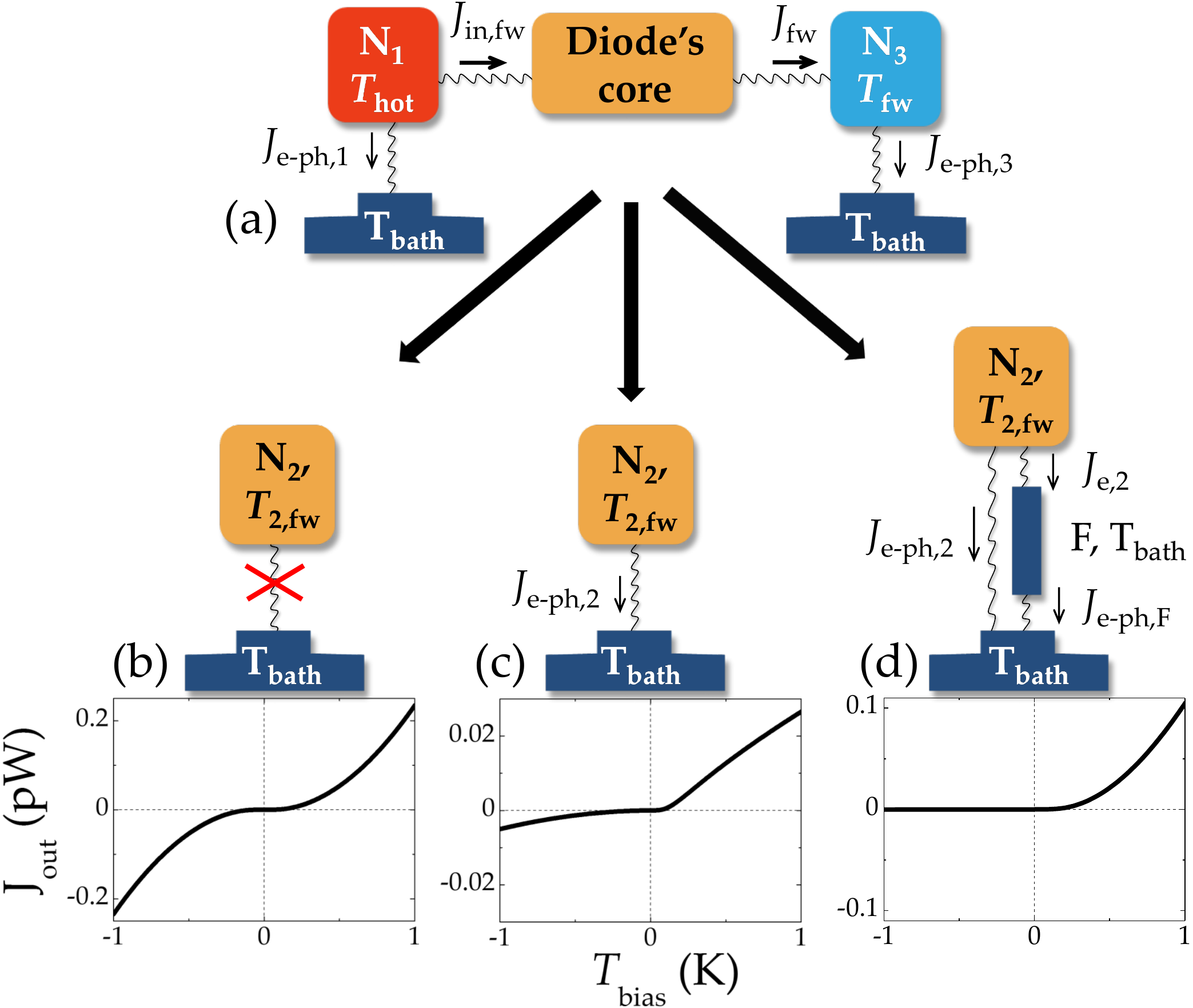}
\caption{(a) Thermal model outlining the relevant heat exchange mechanisms present in the structure. Arrows indicate the heat current directions in the forward temperature bias configuration, which is characterized by $T_{\rm hot}> T_{\rm 2,fw}> T_{\rm fw}>T_{\rm bath}$. Panels (b), (c) and (d) display three possible designs of the diode's core along with the resulting $J_{\rm out}$ vs. $T_{\rm bias}$ characteristics. No rectification is obtained [see panel (b)] unless N$_2$ can release energy to the thermal bath through electron-phonon coupling $J_{\rm e-ph,2}$ [panel (c)] and/or another N electrode labeled F, acting as a thermalizing cold finger [panel (d)].}
\label{Fig2}
\end{figure}
Equations \ref{Qe} and \ref{Qeph} can be used to formulate a thermal model accounting for heat transport through the device. 
The model is sketched in Fig \ref{Fig2}(a) and describes the forward temperature bias configuration, in which the electrodes of the chain reside at temperatures $T_{\rm hot}> T_{\rm 2,fw}> T_{\rm fw}>T_{\rm bath}$. Here, $T_{\rm 2,fw}$ and $T_{\rm fw}$ represent the electronic temperatures of N$_2$ and N$_3$, respectively.
The terms $J_{\rm in,fw}=J_{\rm e}(T_{\rm hot}, T_{\rm 2,fw})$ and $J_{\rm fw}=J_{\rm e}(T_{\rm 2,fw}, T_{\rm fw})$ account for the heat transferred from N$_1$ to N$_3$. The reservoirs can release energy to the phonon bath by means of $J_{\rm e-ph,1}$ and $J_{\rm e-ph,3}$. Photon-mediated thermal transport, \cite{MeschkeNature,Schmidt,Pascal} owing to poor impedence matching, as well as pure phononic heat currents are neglected in our analysis.\cite{GiazottoNature,MartinezNature,MartinezArxiv} We can now write a system of energy-balance equations that account for the detailed thermal budget in N$_2$ and N$_3$ by setting to zero the sum of all the incoming and outgoing heat currents:
\begin{equation}
\left\{
\begin{aligned}
&J_\mathrm{in,fw}(T_\mathrm{hot},T_\mathrm{2,fw})-J_{\rm fw}(T_\mathrm{2,fw},T_{\rm fw})-J_\mathrm{cool}(T_\mathrm{2,fw},T_\mathrm{bath})=0 \\ 
&J_{\rm fw}(T_\mathrm{2,fw},T_{\rm fw})-J_\mathrm{e-ph,3}(T_{\rm fw},T_\mathrm{bath})=0\label{eqs}.
\end{aligned}
\right. 
\end{equation}
Here $J_{\rm cool}(T_\mathrm{2,fw},T_\mathrm{bath})$ is the heat current that flows from N$_2$ to the phonon bath. Since the rectification effectiveness is defined under the condition of equal temperature bias in both the configurations, in Eqs. \ref{eqs} we set $T_{\rm hot}$ and $T_{\rm bath}$ as independent variables and we calculate the resulting $T_{\rm 2,fw}$ and $T_{\rm fw}$. Another system of energy-balance equations can be written and solved for the reverse configuration,\cite{reverse} in which N$_2$ and N$_1$ reach electronic temperatures $T_{\rm 2,rev}$ and $T_{\rm rev}$, respectively. Finally, we can extract the values of $\mathcal{R}$ and $\eta$. 

We discuss in the following the crucial role of N$_2$, which is the core of the proposed thermal diode. Figures \ref{Fig2}(b), \ref{Fig2}(c) and \ref{Fig2}(d) display three possible designs of the central electrode. It is illustrative to start with the simplest one, at least conceptually, consisting of an island perfectly isolated from the phonon bath [see Fig. \ref{Fig2}(b)]. In this case, the term $J_{\rm cool}$ is null and from the energy-balance equations we obtain:
\begin{align}
T_{\rm 2,fw}&=\sqrt{\frac{r}{r+1}T_\mathrm{hot}^2+\frac{1}{r+1}T_{\mathrm{fw}}^2},\\
T_{\rm 2,rev}&=\sqrt{\frac{1}{r+1}T_\mathrm{hot}^2+\frac{r}{r+1}T_{\mathrm{rev}}^2}.
\end{align}
Then, since by assumption we set $\Sigma_1=\Sigma_3$ and $\mathcal{V}_1=\mathcal{V}_3$, we have $T_{\rm fw}=T_{\rm rev}=T_{\rm cold}$ and we can easily show that the rectification efficiency becomes:
\begin{equation}
 \mathcal{R}=\frac{J_{\rm fw}}{J_{\rm rev}}=\frac{(T_{\rm 2,fw}^2-T_\mathrm{cold}^2)}{r(T_{\rm 2,rev}^2-T_\mathrm{cold}^2)}=1,
 \end{equation}
for \emph{every} value of $r$. This means that in this system no rectification occurs regardless the asymmetry of the coupling between N$_2$ and the reservoirs. As a confirmation of the latter result, the lower panel of Fig. \ref{Fig2}(b) displays the symmetrical behavior of $J_{\rm out}$ vs. $T_{\rm bias}$, calculated for $r=$ 100.  

In order to envision a device exhibiting a sizable thermal rectification, it is useful to rewrite $\mathcal{R}\gg 1$ as the following condition for temperatures:
\begin{equation}
\mathcal{R}=\frac{\overline{T}_{\rm fw} \delta T_{\rm fw}}{r \overline{T}_{\rm rev}\delta T_{\rm rev}}\gg 1,\label{rect}
\end{equation}
where $\delta T_{\rm fw (rev)}= T_{\rm 2, fw (rev)}-T_{\rm fw (rev)}$ and the mean temperatures $\overline{T}_{\rm fw (rev)}=(T_{\rm 2,fw (rev)}+T_{\rm fw (rev)})/2$. Expression \ref{rect} indicates a simple approach to pursue our goal: in the reverse configuration the electronic temperatures of N$_1$ and N$_2$ must be similar and close to the lowest temperature in the system, i.e., $T_{\rm bath}$. This condition can be satisfied by setting $r>1$ and by coupling N$_2$ to the phonon bath. In this way, $R_2>R_1$ reduces heat transfer to N$_2$, which is able to release energy to the bath, thereby lowering both $\overline{T}_{\rm rev}$ and $\delta T_{\rm rev}$. On the other hand, the coupling between N$_2$ and the phonon bath should have a limited impact on heat transport in the forward configuration.

Figures \ref{Fig2}(c) and \ref{Fig2}(d) show two possible designs that allow to create a thermal link between N$_2$ and the phonon bath. The former just exploits the natural coupling between electrons and lattice phonons in metals, while the latter requires an additional N electrode labeled F, acting as a thermalizing cold finger. To this end, F must be tunnel-coupled to N$_2$ (with a resistance $R_{\rm F}$) and must reside at $T_{\rm bath}$. We shall demonstrate that the second option provides the best rectification performance, as shown by the dependence of $J_{\rm out}$ on $T_{\rm bias}$ in the lower panels of Fig. \ref{Fig2}(c) and \ref{Fig2}(d). The former curve is calculated for $r=100$ and $\mathcal{V}_2=1\times 10^{-18}$ m$^{-3}$, while the latter is obtained for $r=100$, $\mathcal{V}_2=1\times 10^{-20}$ m$^{-3}$ and $R_{\rm F}=500 \; \Omega$.
\begin{figure}[t]
\centering
\includegraphics[width=1\columnwidth]{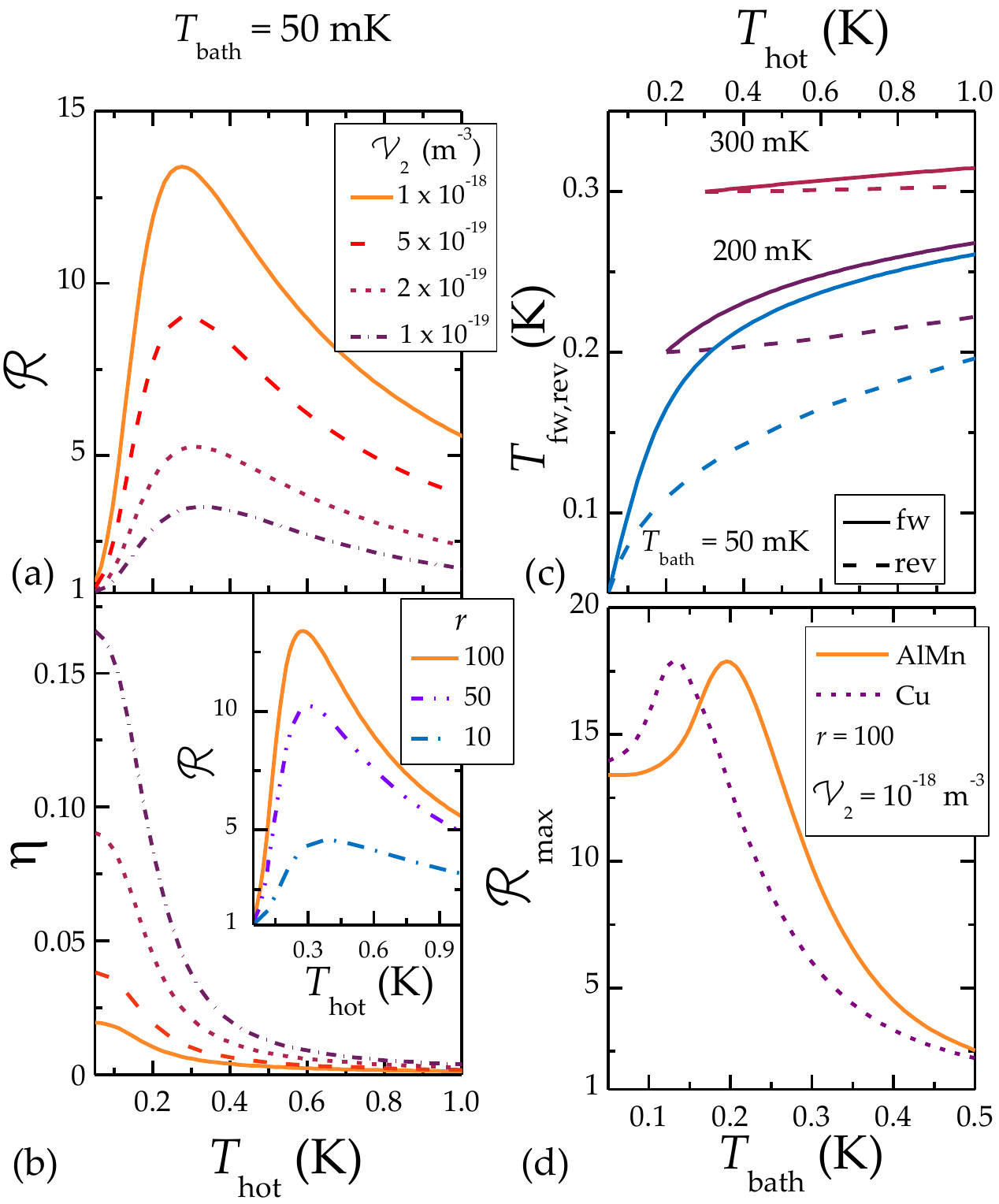}
\caption{Performance of the thermal rectifier with the design sketched in Fig. \ref{Fig2}(c). (a) $\mathcal{R}$ vs. $T_{\rm hot}$ characteristics calculated for different values of $\mathcal{V}_2$ and for $r=100$. (b) Behavior of $\rm \eta$ in the same range of $T_{\rm hot}$ and for the same parameters considered in panel (a). The inset shows $\mathcal{R}$ vs. $T_{\rm hot}$ for $\mathcal{V}_2=1\times 10^{-18}$ m$^{-3}$ and for different values of $r$. In both panels (a) and (b) curves are calculated at $T_{\rm bath}=50$ mK. (c) Diode's output temperatures $T_{\rm fw}$ (solid line) and $T_{\rm rev}$ (dashed lines) vs. $T_{\rm hot}$ at three representative values of $T_{\rm bath}$. (d) $\mathcal{R}_{\rm max}$ vs. $T_{\rm bath}$ for two different N materials.  In panels (c) and (d) we set $\mathcal{V}_2=1\times 10^{-18}$ m$^{-3}$ and $r=$ 100.}
\label{Fig3}
\end{figure}
We focus first on the design sketched in Fig. \ref{Fig2}(c), which corresponds to setting the term $J_\mathrm{cool}=J_{\rm e-ph,2}$ in the energy-balance equations. Figure \ref{Fig3}(a) and \ref{Fig3}(b) show respectively the behavior of $\mathcal{R}$ and $\eta$ vs. $T_{\rm hot}$ in an AlMn device for different values of $\mathcal{V}_2$ at $T_{\rm bath}=50$ mK and for $r=100$. The rectification effectiveness exhibits a non-monotonic trend that reaches a maximum value ($\mathcal{R}_{\rm max}$) around $T_{\rm hot}=300$ mK, whereas $\eta$ has a maximum at $T_{\rm hot}\simeq T_{\rm bath}$ and monotonically approaches zero at larger temperatures. The inset of Fig. \ref{Fig3}(b) displays $\mathcal{R}$ as a function of $T_{\rm hot}$ for different values of the resistance asymmetry $r$ and for $\mathcal{V}_2=1\times 10^{-18}$ m$^{-3}$. It appears evident how increasing $r$ and $\mathcal{V}_2$ generate larger values of $\mathcal{R}$ but gradually suppress the thermal efficiency. Furthermore, the latter parameter clearly indicates that the influence of the electron-phonon coupling increases with temperature (see Eq. \ref{Qeph}), causing strong energy losses in both the configurations. This can be noted also in Fig. \ref{Fig3}(c), which displays the output temperatures $T_{\rm fw}$ and $T_{\rm rev}$ vs. $T_{\rm hot}$ of a diode with $\mathcal{V}_2=1\times 10^{-18}$ m$^{-3}$ and $r=100$ for three representative values of $T_{\rm bath}$. In particular, the derivatives of $T_{\rm fw}$ and $T_{\rm rev}$ strongly decrease as $T_{\rm hot}$ and $T_{\rm bath}$ increase, pinpointing a reduction in the thermal efficiency of the device. Moreover, we also notice that at low $T_{\rm bath}$ the electron-phonon coupling is not much effective in lowering $\overline{T}_{\rm rev}$ and $\delta T_{\rm rev}$. This effect is highlighted in Fig. \ref{Fig3}(d), which compares the $\mathcal{R}_{\rm max}$ dependence on $T_{\rm bath}$ for two devices made of AlMn and Cu. In the AlMn diode $\mathcal{R}_{\rm max}$ reaches a maximum value of $\simeq 18$ at $T_{\rm bath}\simeq 200$ mK, which indicates the temperature where the electron-phonon coupling attains the highest effectiveness in lowering $\overline{T}_{\rm rev} \delta T_{\rm rev}$, while having limited repercussions on $\overline{T}_{\rm fw} \delta T_{\rm fw}$. On the other hand, in Cu $J_{\rm e-ph}$ follows a $T^5$ power law that leads to a shift of the $\mathcal{R}_{\rm max}$ peak to lower values of $T_{\rm bath}$.
\begin{figure}[t]
\centering
\includegraphics[width=1\columnwidth]{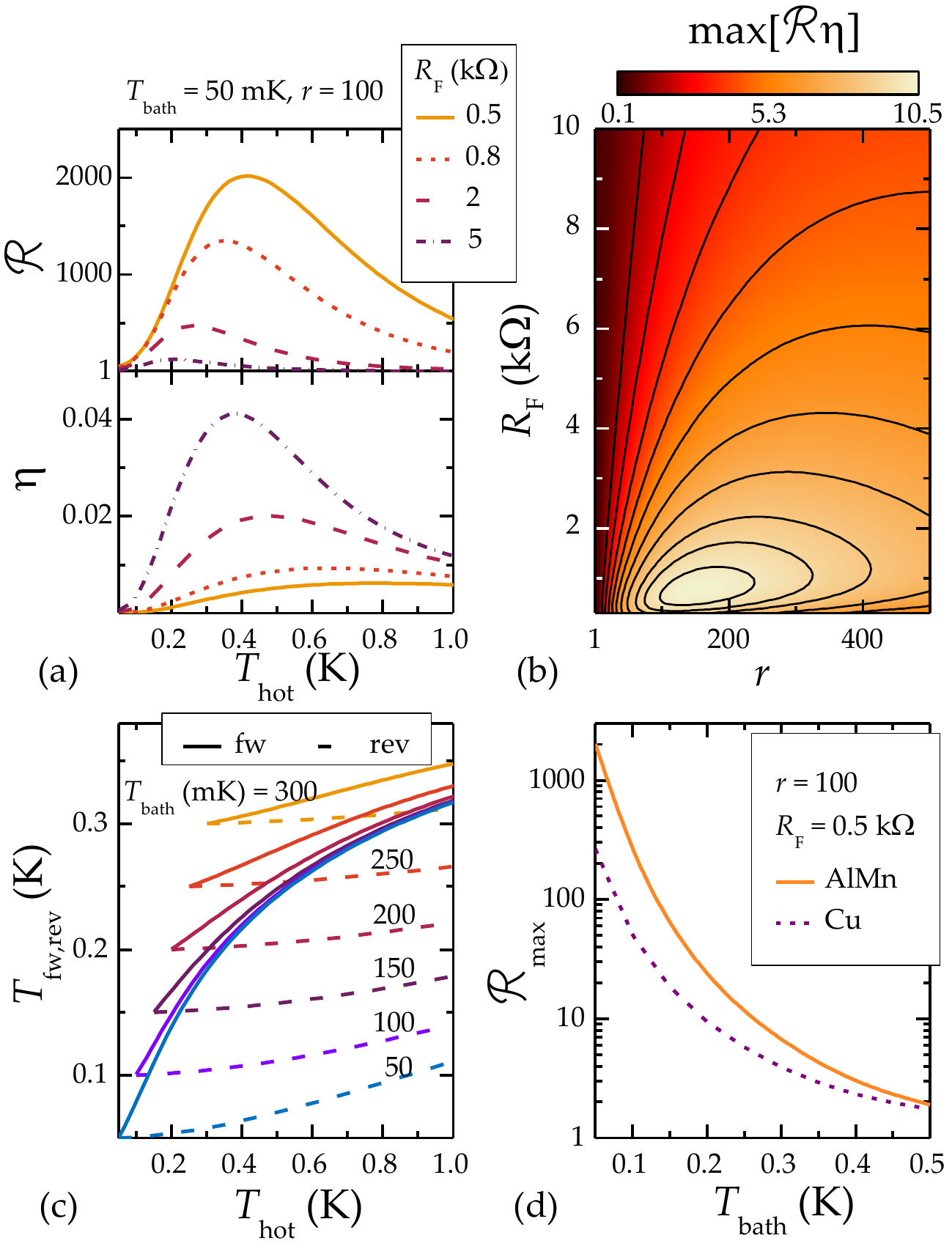}
\caption{Performance of the thermal rectifier with the design sketched in Fig. \ref{Fig2}(d). (a) $\mathcal{R}$ and $\rm \eta$ vs. $T_{\rm hot}$ for different values of the cold finger resistance $R_{\rm F}$. The results are obtained at $T_{\rm bath}=50$ mK and for $r=$ 100. (b) Contour plot showing the maximum value of $\mathcal{R}\eta$ as a function of $R_{\rm F}$ and $r$ at $T_{\rm bath}=50$ mK.  (c) Diode's output temperatures $T_{\rm fw}$ (solid line) and $T_{\rm rev}$ (dashed lines) vs. $T_{\rm hot}$ at different values of $T_{\rm bath}$. (d) $\mathcal{R}_{\rm max}$ vs. $T_{\rm bath}$ for two different N materials. In panels (c) and (d) we set $R_{\rm F}=500$ $\Omega$ and $r=$ 100. All the results have been obtained for $\mathcal{V}_2=1\times 10^{-20}$ m$^{-3}$.}
\label{Fig4}
\end{figure}
These results can be largely improved by tunnel-coupling the electrode F to N$_2$, thereby creating an efficient channel through which the diode's core can release energy. As sketched in Fig. \ref{Fig2}(d), this can be taken into account by setting $J_\mathrm{cool}=J_{\rm e-ph,2}+J_{\rm e,2}$ in the energy-balance equations. Figure \ref{Fig4}(a) shows $\mathcal{R}$ and $\eta$ vs. $T_{\rm hot}$ for different values of $R_{\rm F}$ at $T_{\rm bath}=50$ mK. Remarkably, $\mathcal{R}_{\rm max}\sim 2000$ is obtained for $R_F=500$ $\Omega$, $\mathcal{V}_2=1\times 10^{-20}$ m$^{-3}$ and $r=100$. Similarly to the previous case, increasing $R_{\rm F}$ and $r$ leads to a suppression of the thermal efficiency, which appears to be lower than that obtained above. Nevertheless, $\eta$ presents new important features: it has a non-monotonic behavior with a minimum at $T_{\rm hot}\simeq T_{\rm bath}$ and a peak centered at a specific $T_{\rm hot}$ which depends on the parameters $R_{\rm F}$, $r$ and $\mathcal{V}_2$ and can correspond to high values of $\mathcal{R}$. The low-temperature dependence is dominated by the F channel, which exchanges heat following a $T^2$ dependence. The influence of the cold finger gradually vanishes at larger temperatures and the electron-phonon coupling starts playing the most important role reducing $\eta$. It is then clear how F is highly efficient in keeping $\overline{T}_{\rm rev}$ close to $T_{\rm bath}$ and $\delta T_{\rm rev}$ close to zero, while attenuating its effect on $T_{\rm 2,fw}$ that can reach relatively high temperatures. This mechanism allows us to decrease $\mathcal{V}_2$ and consequently to reduce the influence of the electron-phonon coupling at high $T_{\rm hot}$. By defining the global efficiency of the thermal rectifier $\mathcal{R} \eta$, it is possible to determine the best parameters to optimize the diode's performance. The contour plot shown in Fig. \ref{Fig4}(b) highlights the dependence of the maximum value of $\mathcal{R} \eta$ on $R_{\rm F}$ and $r$, indicating the optimal working region of the proposed device. Figure \ref{Fig4}(c) displays the rectifier's output temperatures $T_{\rm fw}$ and $T_{\rm rev}$ vs. $T_{\rm hot}$ for $r=100$ and $R_{\rm F}=500$ $\Omega$ at different values of $T_{\rm bath}$. The results point out a maximum difference of $\sim 200$ mK between the forward and reverse configurations at $T_{\rm bath}=50$ mK. Finally, Fig. \ref{Fig4}(d) confirms the noxious effect of the electron-phonon coupling on the performance of the diode. As a matter of fact, the behavior of $R_{\rm max}$ as a function of $T_{\rm bath}$ shows that the $T^5$ dependence of $J_{\rm e-ph}$ in Cu is extremely detrimental in the forward configuration and can reduce the rectification effectiveness up to a factor 10.

It is worth mentioning that adding a fourth electrode to the chain would allow us to obtain $\mathcal{R}\neq 1$ even when $r=1$, provided that the coupling between the central islands and the phonon bath is asymmetrical. However, it turns out that this effect cannot improve the performance obtained with the chain made of three electrodes. As a matter of fact, the additional element in the diode's core produces further energy losses and mitigates the temperature gradient $\delta T_{\rm fw}$, thereby reducing $J_{\rm fw}$.
 
In summary, we have proposed and theoretically analyzed the performance of a thermal rectifier consisting of a simple NININ junction. We have demonstrated that a large rectification is achievable if the thermal symmetry of the system is broken and the central electrode is coupled to the phonon bath. Extremely high values of $\mathcal{R}\sim 2000$ can be obtained if a cold finger is connected to the core of the rectifier, thereby creating an efficient channel through which the diode can release energy in the non-transmissive temperature bias configuration. The device could be easily implemented by standard nanofabrication techniques \cite{MartinezArxiv} and, combined with heat current interferometers, \cite{GiazottoNature,MartinezNature} might become one of the building blocks of coherent caloritronic nanocircuits.\cite{GiazottoNature,MartinezNature,MartinezRev} Moreover, its essential design and composition candidate our diode to become a promising tool for thermal management in general-purpose cryogenic electronic applications, even in presence of magnetic fields.

The Marie Curie Initial Training Action (ITN) Q-NET 264034 and the Italian Ministry of Defense through the PNRM project TERASUPER are acknowledged for partial financial support.





\begin{thebibliography}{99}
\bibitem{GiazottoRev} F. Giazotto, T. T. Heikkil\"{a}, A. Luukanen, A. M. Savin, and J. P. Pekola, Rev. Mod. Phys. \textbf{78}, 217 (2006).
\bibitem{Dubi} Y. Dubi and M. Di Ventra, Rev. Mod. Phys. \textbf{83}, 131 (2011).
\bibitem{LiRev} N. Li, J. Ren, L. Wang, G. Zhang, P. H\"{a}nggi, and B. Li, Rev. Mod. Phys. \textbf{84}, 1045 (2012).
\bibitem{BenentiRev} G. Benenti, G. Casati, T. Prosen, and K. Saito, arXiv:1311.4430.
\bibitem{GiazottoNature} F. Giazotto and M. J. Mart\'inez-P\'erez, Nature \textbf{492}, 401 (2012).
\bibitem{MartinezNature} M. J. Mart\'inez-P\'erez and F. Giazotto, Nat. Commun. \textbf{5}, 3579 (2014).
\bibitem{MartinezRev} M. J. Mart\'inez-P\'erez, P. Solinas, and F. Giazotto, J. Low Temp. Phys., 2014, 10.1007/s10909-014-1132-6.
\bibitem{GiazottoTaddei} F. Giazotto, F. Taddei, R. Fazio, and F. Beltram, Appl. Phys. Lett. \textbf{80}, 3784 (2002).
\bibitem{GiazottoHeikkila} F. Giazotto, T. T. Heikkil\"{a}, G. P. Pepe, P. Helist\"{o}, A. Luukanen, and J. P. Pekola, Appl. Phys. Lett. \textbf{92}, 162507 (2008).
\bibitem{NielsenChuang} M. A. Nielsen and I. L. Chuang, Quantum Computation and Quantum Information (Cambridge University Press, 2002).
\bibitem{SpillaArxiv} S. Spilla, F. Hassler, and J. Splettstoesser, arXiv:1311.7561.
\bibitem{Starr} C. Starr, J. Appl. Phys. \textbf{7}, 15 (1936).
\bibitem{RobertsRev} N.A. Roberts and D.G. Walker, Int. J. Therm. Sci., \textbf{50}, 648 (2011).
\bibitem{Segal} L.-A. Wu and D. Segal, Phys. Rev. Lett. \textbf{102}, 095503 (2009).
\bibitem{Segal2} D. Segal, Phys. Rev. Lett. \textbf{100}, 105901 (2008).
\bibitem{Li} B. Li, L. Wang, and G. Casati,  Appl. Phys. Lett. \textbf{88} (2006).
\bibitem{Terraneo} M. Terraneo, M. Peyrard, and G. Casati, Phys. Rev. Lett. \textbf{88}, 094302 (2002).
\bibitem{GiazottoBergeret} F. Giazotto and F. S. Bergeret, Appl. Phys. Lett. \textbf{103}, 242602 (2013).
\bibitem{MartinezAPL} M. J. Mart\'inez-P\'erez and F. Giazotto, Appl. Phys. Lett. \textbf{102}, 182602 (2013).
\bibitem{Ren}J. Ren and J.-X. Zhu, Phys. Rev. B \textbf{87}, 165121 (2013).
\bibitem{Ruokola1}  T. Ruokola and T. Ojanen, Phys. Rev. B \textbf{83}, 241404 (2011).
\bibitem{Kuo} D. M. T. Kuo and Y.C. Chang, Phys. Rev. B \textbf{81}, 205321 (2010).
\bibitem{Ruokola2} T. Ruokola, T. Ojanen, and A.-P. Jauho, Phys. Rev. B  \textbf{79}, 144306 (2009).
\bibitem{Chen} X.-O. Chen, B. Dong, and X.-L. Lei, Chin. Phys. Lett. \textbf{25}, 8 (2008).
\bibitem{BenAbdallah} P. Ben-Abdallah and S.-A. Biehs, Appl. Phys. Lett. \textbf{103}, 191907 (2013).
\bibitem{Chang} C. W. Chang, D. Okawa, A. Majumdar, and A. Zettl, Science \textbf{314}, 1121 (2006).
\bibitem{Kobayashi} W. Kobayashi, Y. Teraoka, and I. Terasaki, Appl. Phys. Lett. \textbf{95}, 171905 (2009).
\bibitem{Tian} H. Tian, D. Xie, Y. Yang, T. L. Ren, G. Zhang, Y. F. Wang, C. J. Zhou, P. G. Peng, L. G. Wang, and L.T. Liu, Sci. Rep. \textbf{2}, 523 (2012).
\bibitem{Scheibner} R. Scheibner, M. K\"{o}nig, D. Reuter, A. D. Wieck, C. Gould, H. Buhmann, and L. W. Molenkamp, New J. Phys. \textbf{10}, 083016 (2008).
\bibitem{MartinezArxiv} M. J.  Mart\'inez-P\'erez, A. Fornieri, and F. Giazotto, arXiv:1403.3052.
\bibitem{Maasilta} L. J. Taskinen and I. J. Maasilta, Appl. Phys. Lett. \textbf{89}, 143511 (2006).
\bibitem{Schmidt} D. R. Schmidt, R. J. Schoelkopf, A. N. Cleland, Phys. Rev. Lett. \textbf{93}, 045901 (2004).
\bibitem{MeschkeNature} M. Meschke, W. Guichard, J. P. Pekola, Nature \textbf{444}, 187 (2006).
\bibitem{Pascal} L. M. A. Pascal, H. Courtois, and F. W. J. Hekking, Phys. Rev. B \textbf{83}, 125113 (2011).
\bibitem{reverse} Energy-balance equations for the reverse configuration can be easily obtained by replacing $J_{\rm in,fw}\Rightarrow J_{\rm in,rev}$, $J_{\rm fw}\Rightarrow J_{\rm rev}$, $J_{\rm e-ph,3}\Rightarrow J_{\rm e-ph,1}$, $T_{\rm 2,fw}\Rightarrow T_{\rm 2,rev}$ and $T_{\rm fw}\Rightarrow T_{\rm rev}$.



\end{thebibliography}
\end{document}